\documentclass[final,3p,times,twocolumn,xcolor=table]{elsarticle}
 \pdfoutput=1 
 
\usepackage{amsmath,amssymb}
\usepackage{graphicx}
\usepackage{multirow}
\usepackage{rotating}
\usepackage{hhline}
\usepackage[active]{srcltx}
\usepackage{colortbl}
\usepackage{floatpag}
\usepackage{mdframed}
\usepackage[nodots,nocompress]{numcompress}

\graphicspath{{./}{./figures/}}

\journal{Archiv}

\begin{document}

\begin{frontmatter}

\title{Semi-automated labelling of medical images: benefits of a collaborative work in the evaluation of prostate cancer in MRI}

\author[label2,label1]{Christian Mata\corref{cor1}}
\ead{christian.mata-miquel@u-bourgogne.fr}
\author[label2]{Alain Lalande}
\author[label2]{Paul Walker}
\author[label1]{Arnau Oliver}
\author[label1]{Joan Mart{\'i}}

\address[label2]{Laboratoire Electronique Informatique et Image (Le2I), UMR CNRS 6306, Universit{\'e} de Bourgogne, Dijon, France.}

\address[label1]{Dept of Computer Architecture and Technology, University of Girona, \\Campus Montilivi, Edifici P-IV, 17071 Girona, Spain}

\cortext[cor1]{Corresponding author. C.~Mata, Facult{\'e} de M{\'e}decine,  Universit{\'e} de Bourgogne, 7, Bld Jeanne d'Arc, BP 87900, 21079 Dijon (France) // Phone: +33380393391.}

\begin{abstract}
\textbf{Purpose}: The goal of this study is to show the advantage of a collaborative work in the annotation and evaluation of prostate cancer tissues from T2-weighted MRI compared to the commonly used double blind evaluation.

\textbf{Methods}: The variability of medical findings focused on the prostate gland (central gland, peripheral and tumoural zones) by two independent experts was firstly evaluated, and secondly compared with a consensus of these two experts. Using a prostate MRI database, experts drew regions of interest (ROIs) corresponding to healthy prostate (peripheral and central zones) and cancer using a semi-automated tool. One of the experts then drew the ROI with knowledge of the other expert's ROI.

\textbf{Results}: The surface area of each ROI as the Hausdorff distance and the Dice coefficient for each contour were evaluated between the different experiments, taking the drawing of the second expert as the reference. The results showed that the significant differences between the two experts became non-significant with a collaborative work.

\textbf{Conclusions}: This study shows that collaborative work with a dedicated tool allows a better consensus between expertise than using a double blind evaluation. Although we show this for prostate cancer evaluation in T2-weighted MRI, the results of this research can be extrapolated to other diseases and kind of medical images.

\end{abstract}


\end{frontmatter}

\section{Introduction}
\label{sec:intro}

The annotation of medical images is subject to an inherent inter-variability between experts, and in some cases, there are also significant differences between the own annotations of the same expert (intra-variability). This difficulty in annotating the medical findings is due to different reasons, including the own images (difficult to understand, low resolution and / or subtle changes, …) the expert that is performing the annotations (experience, tiredness, …) and the working conditions (monitor, annotating device, illuminance, …). It is commonly accepted that a way to reduce the variabilities is by doing the overlap between the annotations performed by different experts that performed them blindly respect to the other experts. In this paper we show that using a collaborative approach, the variabilities between experts can be even more minimised.

\begin{figure*}[t]
\centering
\includegraphics[width=5in]{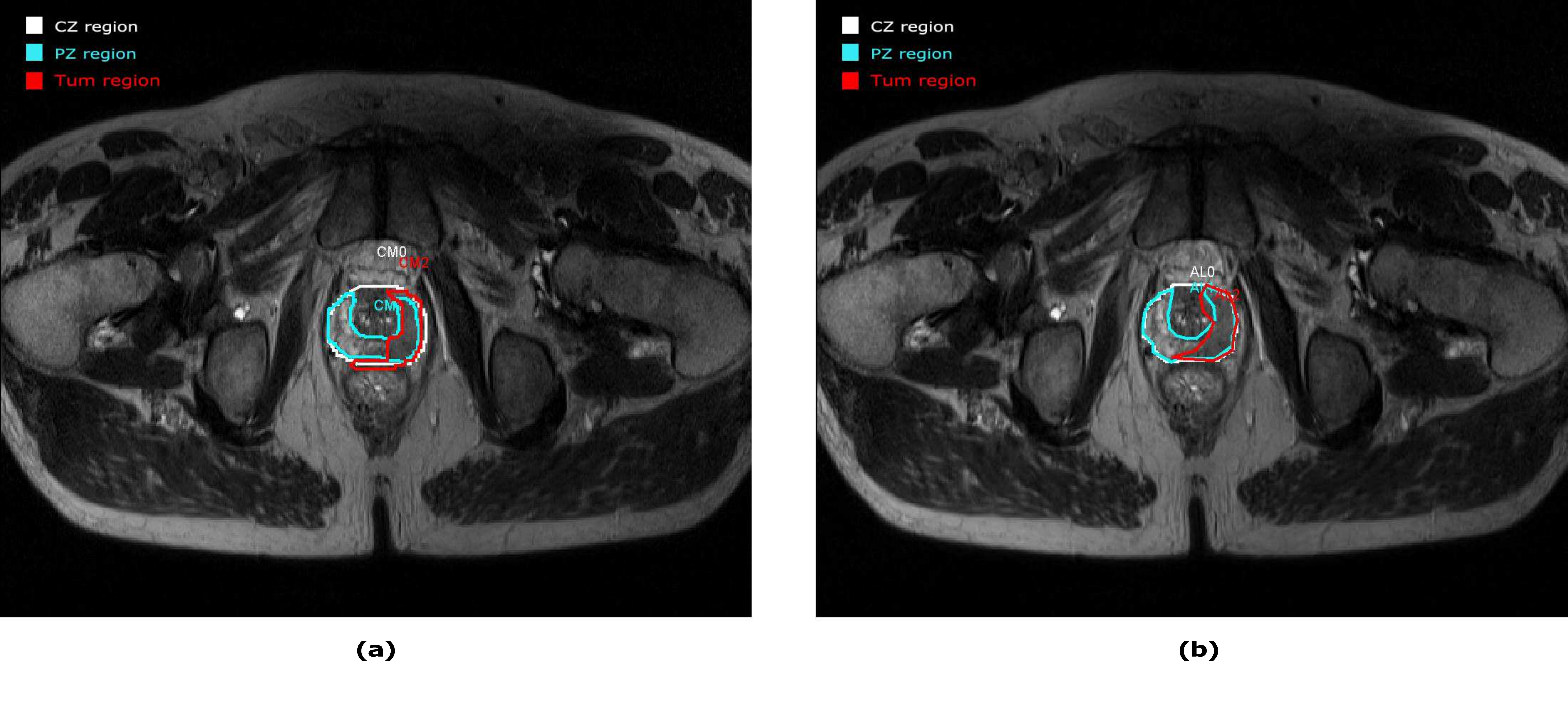}
\caption{Example of a PCa processing from E1 (left image) and from E2 (right image). Notice that the contours are similar for $E1$ and $E2$.}
\label{fig:concordance}
\end{figure*}

We centred our study in the analysis of prostate cancer. Prostate cancer (PCa) remains one of the most commonly diagnosed solid tumours among men. In the United States, there are an estimated 180,890 new cases and 26,120 deaths in 2016~\citep{Siegel14}. Simply speaking the prostate is composed of the peripheral zone (PZ), the transitional zone (TZ) and the central zone (CZ). Most cancer lesions occur in the peripheral zone of the gland. A detailed description of the influence of the prevalence factor risk according the prostate zone is defined in~\citet{Marzo07}. Among the techniques used to detect PCa, Magnetic Resonance Imaging (MRI) allows the non-invasive analysis of the anatomy and the metabolism in the entire prostate gland. MRI has been established as the best imaging modality for the detection, localisation and staging of PCa on account of its high resolution and excellent spontaneous contrast of soft tissues and the possibilities of multi-planar and multi-parameter scanning~\citep{chen08}.

The annotations of the prostate were performed by each expert using the ProstateAnalyzer software~\citep{Mata15}. ProstateAnalyzer allows manual drawing of the different regions of interest (ROI) of the prostate, such as CZ, TZ, PZ and tumour lesion, thanks to the combination of MRI techniques (such as oblique axial T2-weighted, diffusion and perfusion imaging) and magnetic resonance spectroscopy.  The manual drawing of these regions is of crucial importance for the disease prognosis. However, accurate manual annotations need the combination of MRI techniques (such as oblique axial T2-weighted, diffusion and perfusion imaging), turning this task into a challenging work due to the high volume of information present in these images.

The purpose of the present study is to evaluate the variability between experts concerning medical findings in prostate gland regions. The differences observed for the delimitation of the different ROIs between independent evaluation or using a collaborative work by different users is studied. The idea behind this study is to show that collaborative work allows a real consensus between experts and potentially decreases variabilities in their evaluation. Figure 1(a) presents an example of the prostate gland analysis with manual drawing of the TZ (in white), PZ (in blue) and tumour area (in red). We have asked two experts to make these drawings independently on several MR examinations, and as a second step, one expert repeated the drawings with the knowledge of the evaluation of the other expert. Differences in the drawings such as in the volume calculations were compared in order to verify that there is a significant increase of the consensus in the results with collaborative work.

\section{Materials and Methods}

\subsection{Database}
\label{subsec:materials}

A database with prostate MRI based on clinical data with tumour and healthy cases was used. The examinations used in our study contained three-dimensional T2-weighted fast spin-echo (TR / TE / ETL: 3600 ms / 143 ms / 109, slice thickness: $1.25$ mm) images acquired with sub-millimetric pixel resolution in an oblique axial plane. Each study comprises a set of $64$ images.

The annotations were performed using the ProstateAnalyzer tool, that allows the drawing of an annotation in a given MRI modality and automatically it draws the same annotation in the other modalities. We mainly annotate the regions using the T2-weighted image (T2WI). The T2WI modality was chosen because it provides the best depiction of the prostate's zonal anatomy. However, in cases were the tumoural region was better depicted in the other modalities, these were used to help the annotation. The two experts have more than 10 years of experience working with prostate imaging.

\subsection{Evaluation procedure}
\label{subsec:procedure}

The evaluation procedure was performed according to different experiments to analyse the three main ROIs (TZ, PZ and tumour lesion (Tum)). Experts have drawn prostate zones corresponding to PZ, TZ and Tum.

The first experiment (E1) consisted of a prostate study evaluation provided by the first expert. It consisted of drawing ROIs of the prostate gland zones when they are required. For each ROI the surface area value was calculated. Considering the whole images for one patient, the volume of each considered tissue was calculated as the surface of each area multiplied by the slice thickness. Similarly, a second experiment (E2) was carried out independently by a second expert in the same manner as E1. Finally, the first expert repeated the same processing with the knowledge of the drawing and the evaluation performed by the second expert (experiment (E3)). The ProstateAnalyzer drawing tool used offers the possibility of showing all the ROIs drawn by every expert. A minimum delay between (E1) and (E3) should be imposed to prevent the expert from remembering his previous tracing. This was greater than one month in our study. The E2 was considered as the experiment of reference. This means that the comparison procedure to evaluate the influence of the collaborative work was done in two steps: firstly, E1 vs E2 and E3 vs E2, and secondly comparison of these two evaluations.

\begin{figure}[t!]
\centering
\includegraphics[width=2.5in]{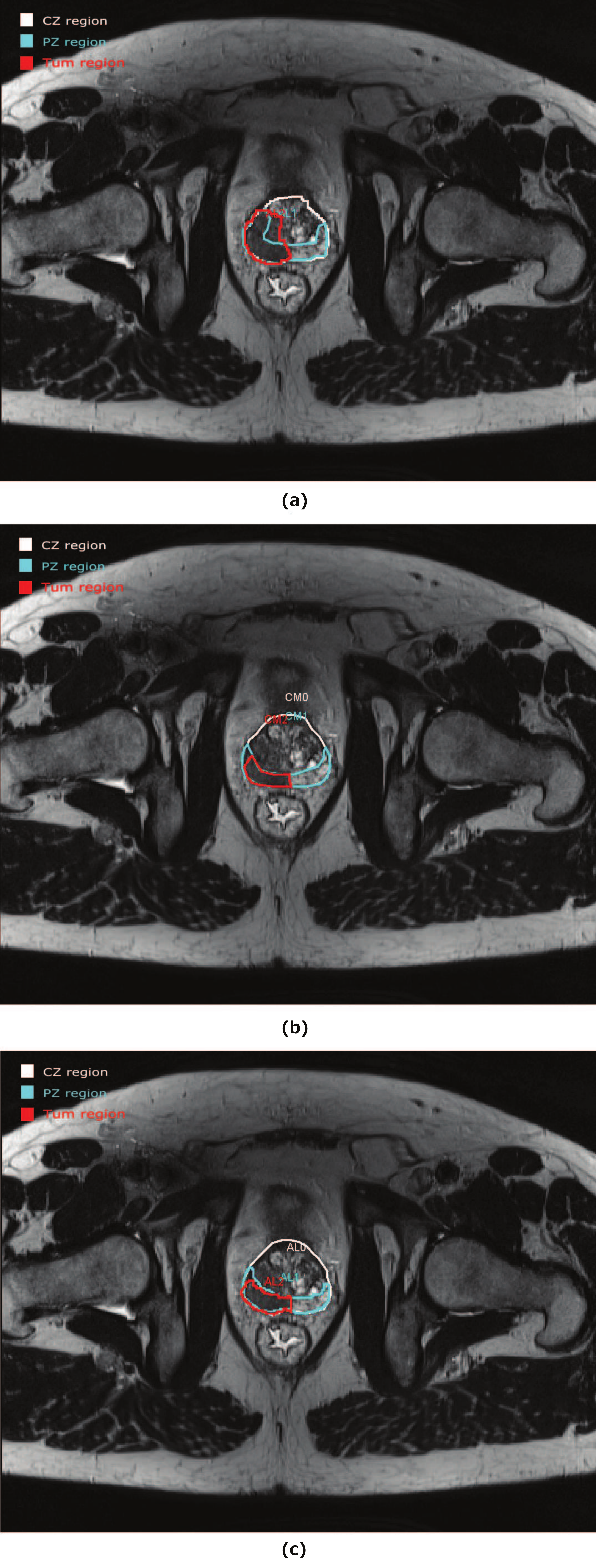}
\caption{Example of a prostate study evaluation from (a) E1 and (b) E2 with a discordance between both drawings for the tumour area. (c) New evaluation of the prostate study from E3 with a good agreement for the tumour area between $E3$ and $E2$.}
\label{fig:discordance}
\end{figure}

\subsection{Evaluation parameters}
\label{subsec:parameters}

The correlation coefficient, the regression analysis and the Bland Altman~\citep{Altman83,Bland86} plot were used to compare the surfaces obtained with $E2$ with those obtained with $E1$ and $E3$, respectively. Moreover, a linear correlation estimation between $E1$ and $E2$, as between $E3$ and $E2$, was performed using a two-sample t-test~\citep{Rice06}. A p-value smaller than $0.05$ was considered as a statistically significant difference. Moreover, the contours obtained from experiment $E2$ were compared with those obtained with $E1$ and also with those obtained with $E3$. An edge-based approach using the Hausdorff distance~\citep{Rote91} and a region-based approach with the Dice index~\citep{Guang04} were considered. The mean and the standard deviation of each parameter for the whole data set were calculated. Again, a two-sample t-test was used to verify if there were any significant difference between the calculation of these parameters taking $E1$ and $E2$, and taking $E3$ and $E2$. Before these analyses, and for each type of tissue, the number of cases in which one expert consider annotating on one image and not the other expert (i.e. corresponding to the upper and lower slices) were counted and presented as a percentage of the total number of processed slices by the second expert.

\begin{table*}[t!]
\centering
\scriptsize
\begin{tabular}{rrrrrrrr}
\multicolumn{1}{l}{}                                                       & \multicolumn{1}{l}{}                                                                                                            & \multicolumn{2}{c}{\cellcolor[HTML]{333333}{\color[HTML]{FFFFFF} CZ}}                                                                                         & \multicolumn{2}{c}{\cellcolor[HTML]{333333}{\color[HTML]{FFFFFF} PZ}}                                                                                         & \multicolumn{2}{c}{\cellcolor[HTML]{333333}{\color[HTML]{FFFFFF} TUM}}                                                                                        \\ \cline{3-8}
\rowcolor[HTML]{9B9B9B}
\multicolumn{1}{l}{\cellcolor[HTML]{333333}{\color[HTML]{FFFFFF} Patient}} & \multicolumn{1}{l|}{\cellcolor[HTML]{333333}{\color[HTML]{FFFFFF} Processed slides}} & \multicolumn{1}{l|}{\cellcolor[HTML]{9B9B9B}{\color[HTML]{FFFFFF} E1 vs. E2}} & \multicolumn{1}{l|}{\cellcolor[HTML]{9B9B9B}{\color[HTML]{FFFFFF} E3 vs. E2}} & \multicolumn{1}{l|}{\cellcolor[HTML]{9B9B9B}{\color[HTML]{FFFFFF} E1 vs. E2}} & \multicolumn{1}{l|}{\cellcolor[HTML]{9B9B9B}{\color[HTML]{FFFFFF} E3 vs. E2}} & \multicolumn{1}{l|}{\cellcolor[HTML]{9B9B9B}{\color[HTML]{FFFFFF} E1 vs. E2}} & \multicolumn{1}{l|}{\cellcolor[HTML]{9B9B9B}{\color[HTML]{FFFFFF} E3 vs. E2}} \\ \hline
\multicolumn{1}{|r|}{Patient 1}                                            & \multicolumn{1}{r|}{18}                                                                                                         & \multicolumn{1}{r|}{6\%}                                                      & \multicolumn{1}{r|}{6\%}                                                      & \multicolumn{1}{r|}{11\%}                                                     & \multicolumn{1}{r|}{0\%}                                                      & \multicolumn{1}{r|}{0\%}                                                      & \multicolumn{1}{r|}{0\%}                                                      \\ \hline
\multicolumn{1}{|r|}{Patient 2}                                            & \multicolumn{1}{r|}{21}                                                                                                         & \multicolumn{1}{r|}{10\%}                                                     & \multicolumn{1}{r|}{10\%}                                                     & \multicolumn{1}{r|}{10\%}                                                     & \multicolumn{1}{r|}{10\%}                                                     & \multicolumn{1}{r|}{0\%}                                                      & \multicolumn{1}{r|}{0\%}                                                      \\ \hline
\multicolumn{1}{|r|}{Patient 3}                                            & \multicolumn{1}{r|}{25}                                                                                                         & \multicolumn{1}{r|}{8\%}                                                      & \multicolumn{1}{r|}{0\%}                                                      & \multicolumn{1}{r|}{8\%}                                                      & \multicolumn{1}{r|}{0\%}                                                      & \multicolumn{1}{r|}{12\%}                                                     & \multicolumn{1}{r|}{4\%}                                                      \\ \hline
\multicolumn{1}{|r|}{Patient 4}                                            & \multicolumn{1}{r|}{30}                                                                                                         & \multicolumn{1}{r|}{7\%}                                                      & \multicolumn{1}{r|}{0\%}                                                      & \multicolumn{1}{r|}{7\%}                                                      & \multicolumn{1}{r|}{0\%}                                                      & \multicolumn{1}{r|}{17\%}                                                     & \multicolumn{1}{r|}{0\%}                                                      \\ \hline
\multicolumn{1}{|r|}{Patient 5}                                            & \multicolumn{1}{r|}{24}                                                                                                         & \multicolumn{1}{r|}{13\%}                                                     & \multicolumn{1}{r|}{0\%}                                                      & \multicolumn{1}{r|}{13\%}                                                     & \multicolumn{1}{r|}{8\%}                                                      & \multicolumn{1}{r|}{13\%}                                                     & \multicolumn{1}{r|}{0\%}                                                      \\ \hline
\multicolumn{1}{|r|}{Patient 6}                                            & \multicolumn{1}{r|}{17}                                                                                                         & \multicolumn{1}{r|}{18\%}                                                     & \multicolumn{1}{r|}{0\%}                                                      & \multicolumn{1}{r|}{12\%}                                                     & \multicolumn{1}{r|}{0\%}                                                      & \multicolumn{1}{r|}{65\%}                                                     & \multicolumn{1}{r|}{0\%}                                                      \\ \hline
\multicolumn{1}{|r|}{Patient 7}                                            & \multicolumn{1}{r|}{26}                                                                                                         & \multicolumn{1}{r|}{12\%}                                                     & \multicolumn{1}{r|}{4\%}                                                      & \multicolumn{1}{r|}{8\%}                                                      & \multicolumn{1}{r|}{0\%}                                                      & \multicolumn{1}{r|}{19\%}                                                     & \multicolumn{1}{r|}{0\%}                                                      \\ \hline
\multicolumn{1}{|r|}{Patient 8}                                            & \multicolumn{1}{r|}{31}                                                                                                         & \multicolumn{1}{r|}{19\%}                                                     & \multicolumn{1}{r|}{10\%}                                                     & \multicolumn{1}{r|}{3\%}                                                      & \multicolumn{1}{r|}{6\%}                                                      & \multicolumn{1}{r|}{0\%}                                                      & \multicolumn{1}{r|}{0\%}                                                      \\ \hline
\multicolumn{1}{|r|}{Patient 9}                                            & \multicolumn{1}{r|}{21}                                                                                                         & \multicolumn{1}{r|}{14\%}                                                     & \multicolumn{1}{r|}{0\%}                                                      & \multicolumn{1}{r|}{14\%}                                                     & \multicolumn{1}{r|}{0\%}                                                      & \multicolumn{1}{r|}{5\%}                                                      & \multicolumn{1}{r|}{0\%}                                                      \\ \hline
\multicolumn{1}{|r|}{Patient 10}                                           & \multicolumn{1}{r|}{25}                                                                                                         & \multicolumn{1}{r|}{16\%}                                                     & \multicolumn{1}{r|}{0\%}                                                      & \multicolumn{1}{r|}{8\%}                                                      & \multicolumn{1}{r|}{0\%}                                                      & \multicolumn{1}{r|}{8\%}                                                      & \multicolumn{1}{r|}{0\%}                                                      \\ \hline
\end{tabular}
 \caption{Counting of the total number of cases for each area (CZ, PZ and Tumour) that have not been evaluated by the two experts between E1 and E2, and E2 and E3.}
 \label{percentagetable}
\end{table*}

\begin{table*}[t!]
\centering
\scriptsize
\begin{tabular}{rrrrrrrrr}
\multicolumn{1}{l}{}                                                     & \multicolumn{2}{c}{\cellcolor[HTML]{333333}{\color[HTML]{FFFFFF} r}}                                                                                          & \multicolumn{2}{c}{\cellcolor[HTML]{333333}{\color[HTML]{FFFFFF} Regression line}}                                                                            & \multicolumn{2}{c}{\cellcolor[HTML]{333333}{\color[HTML]{FFFFFF} Bland-Altman}}                                                                               & \multicolumn{2}{c}{\cellcolor[HTML]{333333}{\color[HTML]{FFFFFF} t-test}}                                                                                     \\ \cline{2-9}
\rowcolor[HTML]{FFFFFF}
\multicolumn{1}{l|}{} & \multicolumn{1}{l|}{\cellcolor[HTML]{9B9B9B}{\color[HTML]{FFFFFF} E1 vs. E2}} & \multicolumn{1}{l|}{\cellcolor[HTML]{9B9B9B}{\color[HTML]{FFFFFF} E3 vs. E2}} & \multicolumn{1}{l|}{\cellcolor[HTML]{9B9B9B}{\color[HTML]{FFFFFF} E1 vs. E2}} & \multicolumn{1}{l|}{\cellcolor[HTML]{9B9B9B}{\color[HTML]{FFFFFF} E3 vs. E2}} & \multicolumn{1}{l|}{\cellcolor[HTML]{9B9B9B}{\color[HTML]{FFFFFF} E1 vs. E2}} & \multicolumn{1}{l|}{\cellcolor[HTML]{9B9B9B}{\color[HTML]{FFFFFF} E3 vs. E2}} & \multicolumn{1}{l|}{\cellcolor[HTML]{9B9B9B}{\color[HTML]{FFFFFF} E1 vs. E2}} & \multicolumn{1}{l|}{\cellcolor[HTML]{9B9B9B}{\color[HTML]{FFFFFF} E3 vs. E2}} \\ \hline
\multicolumn{1}{|r|}{CZ}                                                 & \multicolumn{1}{r|}{0.95}                                                     & \multicolumn{1}{r|}{0.98}                                                     & \multicolumn{1}{r|}{y = 0.9x - 166}                                               & \multicolumn{1}{r|}{y = x - 12}                                                   & \multicolumn{1}{r|}{-261.13 $\pm$ 168.20}                                               & \multicolumn{1}{r|}{-13.07 $\pm$ 118.09}                                                & \multicolumn{1}{r|}{0.01}                                                     & \multicolumn{1}{r|}{0.36}                                                     \\ \hline
\multicolumn{1}{|r|}{PZ}                                                 & \multicolumn{1}{r|}{0.91}                                                     & \multicolumn{1}{r|}{0.94}                                                     & \multicolumn{1}{r|}{y = 0.9x - 96}                                                & \multicolumn{1}{r|}{y = 0.9x + 21}                                                & \multicolumn{1}{r|}{-156.50 $\pm$ 95.71}                                                & \multicolumn{1}{r|}{-10.73 $\pm$ 84.60}                                                 & \multicolumn{1}{r|}{0.01}                                                     & \multicolumn{1}{r|}{0.32}                                                     \\ \hline
\multicolumn{1}{|r|}{TUM}                                                & \multicolumn{1}{r|}{0.96}                                                     & \multicolumn{1}{r|}{0.98}                                                     & \multicolumn{1}{r|}{y = 0.7x - 3}                                                 & \multicolumn{1}{r|}{y = x + 3}                                                    & \multicolumn{1}{r|}{-54.93 $\pm$ 64.34}                                                 & \multicolumn{1}{r|}{-0.08 $\pm$ 27.13}                                                & \multicolumn{1}{r|}{0.02}                                                     & \multicolumn{1}{r|}{0.47}                                                     \\ \hline
\end{tabular}
 \caption{Analysis of CZ, PZ and the tumour (TUM) area calculated (in $mm^{2}$) found in the prostate gland using the surface as the anatomical parameter. $E2$ is the reference and is compared with $E1$ and $E3$.}
 \label{tanatomic}
\end{table*}

\section{Results}
\label{sec:results}

Two examples of a PCa analysis are presented in Figures~\ref{fig:concordance} and~\ref{fig:discordance}. The left image in Figure~\ref{fig:concordance} corresponds to the drawing by the first expert $(E1)$ and in the right image by the second expert $(E2)$. Three ROIs are drawn in both images corresponding to CZ (white), PZ (blue) and Tumour (red). When visually comparing the two drawings, a very good concordance between CZ and PZ areas can be observed. Concerning the tumourous area, a small deviation is found but contours could be considered as relativity close between the two experiments.

However, not all the prostate studies were evaluated with such as good concordance between experiments. An example of discordance is shown in Figure~\ref{fig:discordance}. CZ and PZ have a good correspondence between $E1$ and $E2$ but an important discordance is observed for the tumour area. Then, a new evaluation of a prostate study for $E3$ concerning the tumour area is shown in Figure~\ref{fig:discordance} (c). In this example, we see the real advantage of collaborative work. According to the two experts after adjustments, the tumour area is similar.

\subsection{Anatomic parameters}
\label{subsubsec:anatomicparameters}

Table~\ref{percentagetable} describes for each tissue the number of cases in which one expert drew one area and not the other. It is represented as a percentage of the total number of slices processed by the second expert. Considering all the data sets (then considering all the patients), a percentage of the total cases for both experiments is calculated. For CZ, this percentage is equal to 12\% between $E1$ vs. $E2$ and 3\% between $E3$ vs. $E2$. In the case of PZ, this percentage is equal to 9\% between $E1$ vs. $E2$ and 3\% between $E3$ vs. $E2$.  Finally, for the tumour it is 13\% between $E1$ vs. $E2$ and 0\% between $E3$ vs. $E2$. It can be seen that in the second comparison between $E3$ vs. $E2$ the number of discordance for each area is drastically reduced.

The correlation coefficient $(r)$, regression line, Bland Altman and two-sample t-test calculated for the three zones are presented in Table~\ref{tanatomic}. In general, the results have been improved between $E3$ vs. $E2$ compared with $E1$ vs. $E2$. The Bland-Altman test shows a better agreement between $E3$ vs. $E2$ than $E1$ vs. $E2$, whatever the area. Bland-Altman is also calculated for the volume evaluation. For CZ it is $40\pm17 mm^{3}$ between $E1$ vs. $E2$ and $-0.9\pm3 mm^{3}$ between $E3$ vs. $E2$, for the PZ it is $20\pm13 mm^{3}$ between $E1$ vs. $E2$ and $3\pm12 mm^{3}$ between $E3$ vs. $E2$, and for tumour it is $7\pm6 mm^{3}$ between $E1$ vs. $E2$ and $0.4\pm0.9 mm^{3}$ between $E3$ vs $E2$. According to the two-sample t-test, there is no significant overlap between $E3$ vs. $E2$ whatever the considered area, while there are always significant differences in the results between $E1$ vs. $E2$

\begin{figure*}[t!]
\centering
\includegraphics[width=6in]{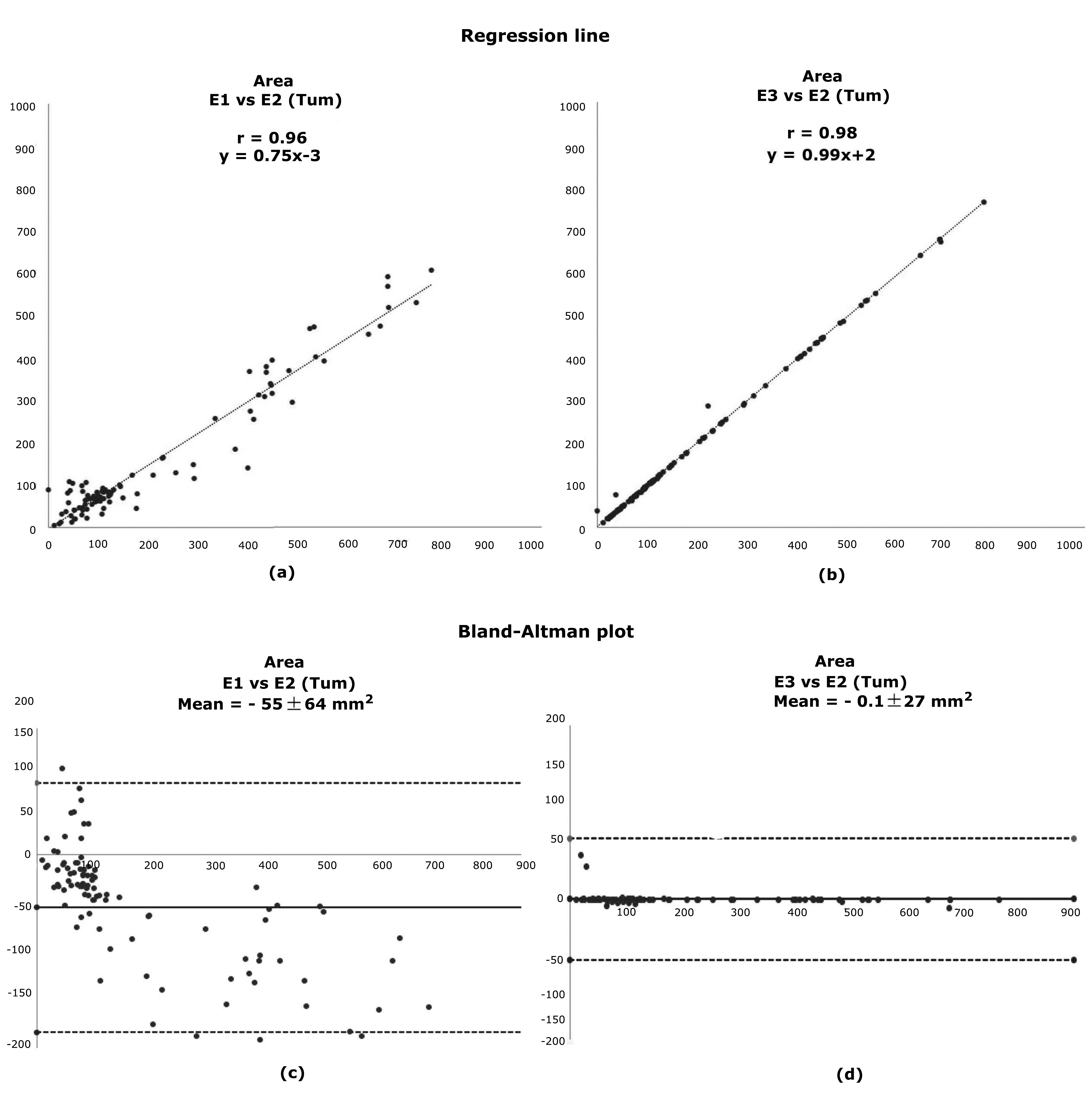}
\caption{Regression analysis obtained between (a) E1 vs. E2 and (b) E2 vs. E3 and the corresponding Bland-Altman plots obtained between (c) E1 vs. E2 and (d) E2 vs. E3 concerning the calculation of the surface of a tumour.}
\label{fig:results}
\end{figure*}

Figure~\ref{fig:results} (a) and (b) illustrate the linear regression analysis concerning the evaluation of the tumour area. The tumour area has been chosen due to its importance and because this area is more difficult to analyze and provides more variations among experts. When comparing the two obtained regression lines, an improvement is noted in Figure~\ref{fig:results}(b)[$E3 vs. $E2 Tum] with a slope of $0.99$ compared with Figure~\ref{fig:results}(a)[$E1 vs. $E2 Tum] with a slope of $0.75$. In the corresponding Bland-Altman plots, in Figure~\ref{fig:results}(d) it can be seen that the mean of the difference between $E3$ vs. $E2$ is close to zero, meaning that there is little bias between the two measurements. Moreover, there is a decrease of the standard deviation.

\subsection{Contour evaluation}
\label{subsec:edgeregionparameters}

The Hausdorff distance and the Dice index between the different drawings are presented in Table~\ref{thausdorff}. Again, between $E3$ vs. $E2$, an improvement is observed with respect to the results obtained between $E1$ vs. $E2$. The mean Hausdorff distance is reduced in all the cases. In the same way, the analysis of the Dice Index is around $0.9$ between $E3$ vs. $E2$ whatever the area, whereas is not this the case for $E1$ vs. $E2$. The differences between $E1$ vs. $E2$ and $E3$ vs. $E2$ are all significant, whatever the considered parameter.

\section{Discussion}
\label{sec:discussion}

In this paper we analyse the ground-truth obtention of a prostate cancer analysis using MRI by two different experts. In contrast of what is currently done for obtaining a ground-truth, where the evaluation of the different experts is independently done and afterwards objectively (or subjectively) merged, in this paper, we study the interest of a collaborative work, where the ground-truth is obtained by the two experts, but with the second one knowing the opinion of the first one. Exhaustive evaluations of the medical findings in different regions of the prostate gland from T2WI for several experiments using our dedicated tool application have been performed. ROIs were drawn in the prostate gland (TZ, PZ and tumour area if present) on images from a prostate MRI database. We proved that using the collaborative approach, the variability between the experts is significantly reduced.

\begin{table*}[t!]
\centering
\scriptsize
\begin{tabular}{l|r|r|r|r|}
\cline{2-5}
& \multicolumn{2}{c|}{\cellcolor[HTML]{333333}{\color[HTML]{FFFFFF} Hausdorff distance}}                                                                        & \multicolumn{2}{c|}{\cellcolor[HTML]{333333}{\color[HTML]{FFFFFF} Dice Index}}                                                                                \\
\rowcolor[HTML]{FFFFFF}
\multicolumn{1}{l}{} & \multicolumn{1}{|l|}{\cellcolor[HTML]{9B9B9B}{\color[HTML]{FFFFFF} E1 vs. E2}} & \multicolumn{1}{c|}{\cellcolor[HTML]{9B9B9B}{\color[HTML]{FFFFFF} E3 vs. E2}} & \multicolumn{1}{c|}{\cellcolor[HTML]{9B9B9B}{\color[HTML]{FFFFFF} E1 vs. E2}} & \multicolumn{1}{c|}{\cellcolor[HTML]{9B9B9B}{\color[HTML]{FFFFFF} E3 vs. E2}} \\ \hline
\multicolumn{1}{|l|}{CZ}                                                   & 8 $\pm$ 3                                                                        & 4 $\pm$ 1                                                                        & 0.7 $\pm$ 0.2                                                                    & 0.9 $\pm$ 0.1                                                                    \\ \hline
\multicolumn{1}{|l|}{PZ}                                                   & 11 $\pm$ 5                                                                       & 5 $\pm$ 2                                                                        & 0.6 $\pm$ 0.2                                                                    & 0.9 $\pm$ 0.1                                                                    \\ \hline
\multicolumn{1}{|l|}{TUM}                                                  & 10 $\pm$ 4                                                                       & 8 $\pm$ 11                                                                       & 0.7 $\pm$ 0.1                                                                    & 0.9 $\pm$ 0.1                                                                           \\ \hline
\end{tabular}
 \caption{Analyses of Hausdorff distance (in $mm^{2}$) and Dice index for the CZ, PZ and tumour area (TUM). A $p$ of $<$ $0.05$ between $E1$ vs $E2$ and $E3$ vs $E2$ is found in all the cases.}
 \label{thausdorff}
\end{table*}

Even if the results are expected, this study shows that evaluation of medical examinations with a knowledge of the other expert reduces drastically the differences between processing. In particular, significant differences between the two experts become non-significant when there is a collaborative work. We probably cannot conclude that the diagnosis was improved, but we will agree that there is a consensus between experts. In general, this involves an increase in the quality of the diagnosis. An alternative point of view would be to affirm that the experiment E3 is biased. An additional experiment could be performed: the second expert could repeat the process knowing what did the first one. However, in our opinion, this is not necessary because the main objective of this study is to objectively show that collaborative work in current clinical practice can provide a real consensus between experts even if there is potentially a bias in the evaluation process.

In a previous work we also analysed the state of the art in automatic prostate segmentation~\citep{Ghose12}. In that work we drew a set of open problems mainly related with the evaluation procedure. This can be summarised as (1) variabilities in the ground-truth, (2) unavailability of public prostate datasets, and (3) lack os standardised metrics for evaluation. We believe the work presented in this paper presents the roots for designing a proper dataset for prostate evaluation. The collaborative work explained in this study is the first step for obtaining a reliable ground-truth, without expert variabilities, which automatic algorithms could be robustly compared.

As a conclusion, although collaborative work requires more time, it allows the improvement of the management of patients with prostate cancer by providing consensual diagnosis, in particular in complex cases.

\footnotesize{
\section*{Acknowledgements}
This work was partially funded by the Spanish R+D+I grant n. TIN2012-37171-C02-01, by UdG grant MPCUdG2016/022. The Regional Council of Burgundy under the PARI 1 scheme also sponsored this work. C. Mata held a Mediterranean Office for Youth mobility grant.
}

\bibliographystyle{model3-num-names}

\end{document}